\documentclass[12pt]{article}
\usepackage{amsmath}
\usepackage{amssymb}
\usepackage{graphicx}
\begin{document}
\setcounter{page}{1}
\def\theequation{\arabic{section}.\arabic{equation}}
\def\theequation{\thesection.\arabic{equation}}
\setcounter{section}{0}

\title{Verification of CPT--invariance of QED bound states for
the production of muonium or antimuonium in scattering of electrons or
positrons by nuclei}

\author{E. A. Choban\thanks{E--mail: choban@part.hop.stu.neva.ru}
\ and G. A. Kazakov}

\date{\today}

\maketitle

\begin{center}
{\it Department of Theoretical Physics, State Technical University of
St. Petersburg,\\ 195251 St. Petersburg, Russian Federation}
\end{center}

\begin{center}
\begin{abstract}
A possibility of a verification of CPT--invariance of QED for bound
states by example of muonium or antimuonium produced in reactions of
scattering of electrons or positrons by nuclei is considered. The
number of events of the muonium production is estimated for
contemporary accelerators. The method of the detection of muonium by
measuring of oscillations of the decay curve caused by the
interference between the ground and excited state of muonium is
suggested. The admixture of the excited muonium to the final state is
calculated.
\end{abstract}
\end{center}

\newpage

\section{Introduction}
\setcounter{equation}{0}

\hspace{0.2in} The verification of CPT--invariance of a quantum
field theory (QFT) (in particular, Quantum Electrodynamics (QED))
is a meaningful problem of high energy physics, since the
postulates of QFT are locality and relativistic invariance [1,2].
As has been shown in Ref.[3], this leads to invariance of the
Lagrangian with respect to C--, P-- and T--invariance. The
simplest consequence CPT--invariance, the equality of masses of a
particle and its antiparticle, is verified at present with a great
accuracy. For example, for the masses of $\mu^+$ and $\mu^-$ one
has $m_{\mu^+}/m_{\mu^-} = 1.000024 \pm 0.000078$ [4]. However,
the problem of CPT--invariance of bound states is still much less
clarified. For example, in Ref.[5] there has been suggested to
treat the production of the antihydrogen for $\bar{p}Z$ collisions
with a subsequent analysis of Lamb shifts $2S_{1/2} - 2P_{1/2}$ of
the transitions between hydrogen and antihydrogen. However,
nowadays the available statistics of events is n   ot enough to
make a definite conclusion. It is smaller compared with the
required by a factor 40.

\section{Cross section for reactions (1)}
\setcounter{equation}{0}

\begin{figure}[bthp]
\centering\includegraphics[width=340pt]{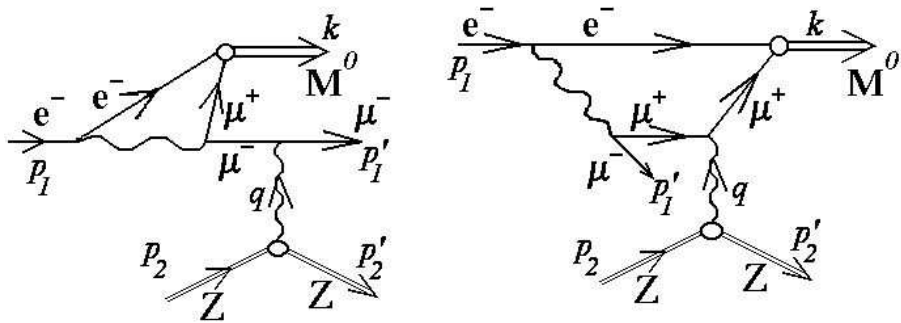}
\caption{Feynman diagrams giving the dominant contribution to the
amplitude of reactions (\ref{label2.1})} \label{f:1}
\end{figure}

\hspace{0.2in} In this paper we suggest to treat the production of a
muonium $M^0$ or antimuonium $\bar{M}^0$ in reactions
\begin{eqnarray}\label{label2.1}
e^- + Z \to Z + M^0 + \mu^- \quad,\quad e^+ + Z \to Z + \bar{M}^0 +
\mu^+,
\end{eqnarray}
where $M^0(\bar{M}^0)$ is a bound state of $\mu^+$, $e^-$
($\mu^-$, $e^+$). The dominant contribution to the amplitude of
reactions (\ref{label2.1}) comes from the diagrams depicted in
Fig.\ref{f:1}.

\begin{figure}
 \centering\includegraphics[width=310pt]{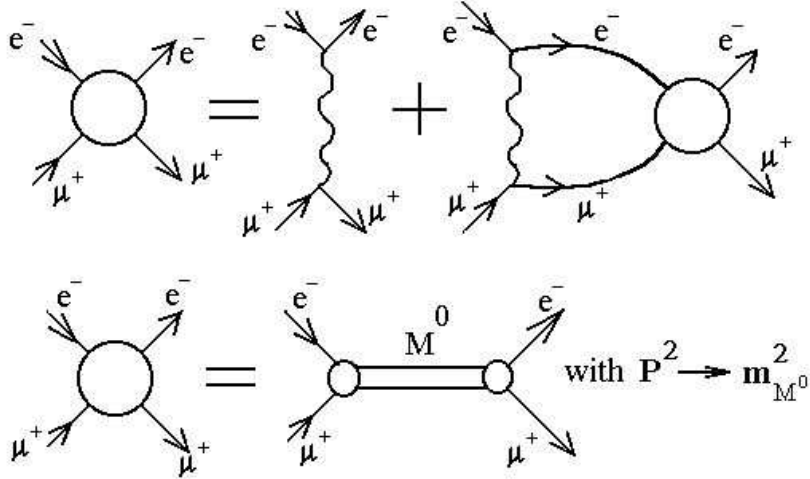}
 \caption{The Bethe--Salpeter equation in the diagrammatic form.}
 \label{f:2}
\end{figure}

Before we have written the amplitude of the reactions
(\ref{label2.1})(we would make everything by example of the
reaction $e^-Z\to ZM^0\mu^-$) we suggest to discuss the vertex of
recombination $e^- + \mu^+ \to M^0$, $e^+ + \mu^- \to \bar{M}^0$.
This vertex has been derived in [6] within the Bethe--Salpeter
equation with a kernel approximated by the one--photon exchange.
The set diagrams depicting this kernel is given in
Fig.\ref{f:2}\,\footnote{In Ref.[7] instead of Fig.\ref{f:2} there has been
occasionally adduced the diagrams in Fig.\ref{f:1} of the present paper.}.

If for the diagrams in Fig.\ref{f:1} to denote the product of the Green
function of electron $\hat{G}_{e^-}(p) = 1/(\hat{p} - m_e)$, the
vertex of recombination $e^- + \mu^+ \to M^0$, $\hat{\Gamma}_0$, and
the Green function of the $\mu^+$--meson $\hat{G}_{\mu^+}(k - p) =
1/(\hat{p} - \hat{k} - m_{\mu})$ as the quantity
\begin{eqnarray}\label{label2.2}
\hat{X}(p, k - p) = \hat{G}_{\mu^+}(k -
p)\,\hat{\Gamma}_0\,\hat{G}_{e^-}(p),
\end{eqnarray}
the solution of the Bethe--Salpeter equation for $\hat{X}(p, k - p)$
one can represent in the form [6]
\begin{eqnarray}\label{label2.3}
\hat{X}(p, k - p)
=(1-\gamma_0)\gamma_5\frac{\sqrt{m_{\mu}}}{2\,i}\frac{\displaystyle
\Big(\frac{\vec{p}^{\,2}}{2m_e} -
E_{bound}\Big)\,\Phi(\vec{p}\,)}{\displaystyle \Big(p_0 - m_e -
\frac{\vec{p}^{\,2}}{2m_e} + i\,0\Big)\Big(p_0 - k_0 + m_{\mu} +
\frac{\vec{p}^{\,2}}{2m_{\mu}} - i\,0\Big)},
\end{eqnarray}
such as the Fourier transform of the wave function of the electron in
$M^0$ reads
\begin{eqnarray*}
\Phi(\vec{p}\,) = \frac{8\sqrt{\pi a^3_0}}{(1 + p^2\,a^2_0)^2},
\end{eqnarray*}
where $a_0 = 1/m_e\alpha$ with $\alpha = e^2/4\pi$ is the fine
structure constant defined in units $\hbar = c = 1$.

Taking into account the diagrams in Fig.\ref{f:1} and using the vertex
(\ref{label2.3}), we obtain the amplitude of the reaction
(\ref{label2.1}) with $e^-$ in the following form
\begin{eqnarray}\label{label2.4}
\hspace{-0.3in}&&M =
\frac{(4\pi\alpha)^2}{q^2}\,\Psi(0)\,{\ell}_{\nu}\,
\frac{\sqrt{m_{\mu}}}{m_e\,(\omega^2 - 2q\cdot k)^2(q^2 - 2q\cdot
k)}\,\bar{u}(p\,'_1)\,\gamma_5\,\Big\{2(2q\cdot k -
\omega^2)\,(q\cdot p\,'_1\,\gamma_{\mu}\nonumber\\
\hspace{-0.3in}&&~~~ - \hat{q}\,p\,'_{1\mu}) + m_{\mu}\,[(q^2 -
2q\cdot k)\,(2\,p\,'_{1\mu} - \gamma_{\mu}\hat{q}) - (\omega^2 -
2q\cdot k)\,(\hat{q}\,\gamma_{\mu} - 2\,k_{\mu})]\Big\}\,u(p_1),
\end{eqnarray}
where $m_e$ and $m_{\mu}$ are masses of the electron and $\mu$--meson,
respectively, $\omega^2 = (p\,'_1 + k)^2$ is the squared invariant
mass of $(M^0,\mu^-)$ system, ${\ell}_{\mu}$ is the electromagnetic
current of the nucleus, and $\Psi(0) = 1/\sqrt{\pi a^3_0}$.

The squared amplitude would involve the tensor $R_{\mu\nu}$ describing
the lower block of the diagrams in Fig.\ref{f:1} containing the
electromagnetic form factors of the nucleus $F_{1Z}$ and $F_{2Z}$:
\begin{eqnarray}\label{label2.5}
\hspace{-0.3in}&&R_{\mu\nu} = 4\Big\{F^2_{1Z}(q^2)\,
\Big[2\,p_{2\mu}p_{2\nu} - (p_{2\mu}q_{\nu} + p_{2\nu}q_{\mu}) +
\frac{1}{2}\,q^2\,g_{\mu\nu}\Big]\nonumber\\
\hspace{-0.3in}&&+F^2_{2Z}(q^2)\,\Big[2\,q^2\,m^2_Z\,g_{\mu\nu} +
q^2\,(p_{2\mu}q_{\nu} + p_{2\nu}q_{\mu}) -
q_{\mu}q_{\nu}\,\Big(\frac{1}{2}\,q^2 + 2\,m^2_Z\Big) -
2\,q^2\,p_{2\mu}p_{2\nu}\Big]\Big\}.
\end{eqnarray}
Let us introduce $x = \cos\theta$, where $\theta$ is a polar angle of
the momentum of $M^0$ defined in the center of mass frame (CMF) of the
system $(M^0,\mu^-)$ relative to the momentum of the electron in the
CMF $(e^-,Z)$. Using Eqs.(\ref{label2.4}) and (\ref{label2.5}) one can
easily get the differential cross section with respect to $\omega^2$
and $x$ which reads
\begin{eqnarray}\label{label2.6}
\frac{d^2\sigma}{d\omega^2dx} &=&
4\,F^2_{1Z}(0)\,|\Psi(0)|^2\pi\,\alpha^4\,\frac{m_{\mu}}{m^2_e}\,\sqrt{1
- \frac{4m^2_{\mu}}{\omega^2}}\,{\ell
n}\Big(\frac{s^3}{m^2_Z\omega^4}\Big)\,\frac{1}{\displaystyle 1 +
x\,\sqrt{1 - \frac{4m^2_{\mu}}{\omega^2}}}\nonumber\\
&&\times\,\frac{s - \omega^2}{s\,\omega^6}\left\{1 +
\frac{\displaystyle 8m^2_{\mu}\sqrt{1 -
\frac{4m^2_{\mu}}{\omega^2}}\Big[(1-x^2)\sqrt{1 -
\frac{4m^2_{\mu}}{\omega^2}} + 2\,x\Big]}{\displaystyle \omega^2\Big[1
- x^2\Big(1 - \frac{4m^2_{\mu}}{\omega^2}\Big)\Big]\,\Big(1 -
x\,\sqrt{1 - \frac{4m^2_{\mu}}{\omega^2}}\Big)^2}\right\},
\end{eqnarray}
where $F_{1Z}(0) = Z$. The expression (\ref{label2.6}) is calculated
within the Weizs\"acker--Williams approach. One can see from
Eq.(\ref{label2.6}) that in the CMF ($M^0,\mu^-$) the muonium produces
itself mainly backward. After the integration of (\ref{label2.6}) over
$x$ one obtains the differential cross section with respect to
$\omega^2$:
\begin{eqnarray}\label{label2.7}
\hspace{-0.3in}&&\frac{d\sigma}{d\omega^2} =
2\,Z^2\,\alpha^7\,\Big(\frac{m_e}{m_{\mu}}\Big)\,\sqrt{1 -
\frac{4m^2_{\mu}}{\omega^2}}\,{\ell
n}\Big(\frac{s^3}{m^2_Z\omega^4}\Big)\,\frac{s-\omega^2}{s\omega^4}\Bigg\{1
-
12\,\frac{m^2_{\mu}}{\omega^2}\nonumber\\
\hspace{-0.3in}&& +
\left[4\,\left(\frac{m^2_{\mu}}{\omega^2}\right)^2\left(1 -
6\,\frac{m^2_{\mu}}{\omega^2}\right) +
2\,\frac{m^2_{\mu}}{\omega^2}\right]\,\frac{1}{\displaystyle \sqrt{1 -
4m^2_{\mu}/\omega^2}}\,{\ell n}\frac{1 + \sqrt{1 -
4m^2_{\mu}/\omega^2}}{1 - \sqrt{1 - 4m^2_{\mu}/\omega^2}}\Bigg\}.
\end{eqnarray}
From Eq.(\ref{label2.7}) one gets that the maximum of the cross
section is located in the vicinity of the threshold of the
$(M^0,\mu^-)$--system production, $\omega^2 \sim 4\,m^2_{\mu}$, and
the quantity $d\sigma/d\omega^2$ falls substantially with $\omega^2$
and behaves like $d\sigma/d\omega^2 \sim 1/\omega^4$.

Let us now obtain the total cross section of the reactions
(\ref{label2.1}). For this aim we introduce a variable $\xi =
4m^2_{\mu}/\omega^2$ and integrate (\ref{label2.7}) over $\xi$. This
yields
\begin{eqnarray}\label{label2.8}
\hspace{-0.5in}&&\sigma =
\Big(\frac{Z^2\alpha^7m_e}{16\,m^3_{\mu}}\Big)\int\limits^1_0f(\xi)
\,d\xi,\nonumber\\ \hspace{-0.5in}&&f(\xi) = {\ell
n}\Bigg(\frac{s^3\xi^2}{16m^2_Zm^4_{\mu}}\Bigg)\,\left\{2\,(4
-3\,\xi^2)\sqrt{1-\xi} + \xi\,(4 +2\,\xi - \xi^2)\,{\ell
n}\left(\frac{1 + \sqrt{1-\xi}}{1 - \sqrt{1-\xi}}\right)\right\}\!.
\end{eqnarray}
Let us estimate the values of the cross sections for the reactions
(\ref{label2.1}) for Tevatron--DIS (FNAL) [8] and LHC [9]
(we mean the beams of daughter leptons). The
cross section is proportional to $Z^2$. Hence, it can be enhanced
by using the target with a big $Z$ [5], for example, Radon
${^{220}}{\rm Rn}_{86}$ having a spin $1/2$. The values of the
cross sections, different luminosities and an expected number
of events for one year are adduced in Table.

\vspace{0.2in}

\noindent Table. The data for the cross sections for the reactions
(\ref{label2.1}) and the expected number of events for one year.
\vspace{0.1in}

\hspace{-0.2in}\begin{tabular}{|c|c|c|c|c| }\hline \cline{1-5}
Accelerator & $\sqrt{s}$, GeV &$\sigma$, fb& L, ${\rm
cm^{-2}\,s^{-1}}$ & N \\ \hline FNAL (Tevatron--DIS) & 477 & 17 &
$2.1\times 10^{32}$ & $1.1\times 10^2$ \\ \hline LHC & 14000 & 28
& $10^{33} - 10^{34}$ & $8.8\times 10^2 - 8.8\times 10^3$ \\
\hline
\end{tabular}

\section{Method of detection and estimate of  admixture of
excited states of muonium and antimuonim}
\setcounter{equation}{0}

\hspace{0.2in} Let us consider the method of the detection of the
muonium $M^0$ and antimuonium $\bar{M}^0$. For the reactions
(\ref{label2.1}) $M^0$ and $\bar{M}^0$ would be production both in the
ground state and in the excited one as well
\begin{eqnarray}\label{label3.1}
\Psi_{M^0}(0) = C_1\Psi_1(0) + C_2\Psi_2(0).
\end{eqnarray}
Emphasize that the solution of the Bethe--Salpeter equation
(\ref{label2.3}) has been obtained for the production of the muonium
in the ground state. The mechanisms of the production of an excited
muonium, from which there follows that the coefficient $C_2$ contains
an additional small parameter, would be discussed below. Taking into
account the decay of muonium, the partial width of which we denote as
$\Gamma_{M^0}$, one can represent the wave function of muonium in the
form
\begin{eqnarray}\label{label3.2}
\Psi_{M^0}(t) = \Big(C_1\Psi_1(t) + C_2\Psi_2(t)\Big)\,\exp\Big( -
\frac{\Gamma_{M^0}t}{2\,\hbar}\Big).
\end{eqnarray}
Here $\Psi_1(t)$ and $\Psi_2(t)$ are defined by
\begin{eqnarray}\label{label3.3}
\Psi_1(t) = \Psi_1\,\exp\Big( - \frac{i}{\hbar}\,E_1t\Big)\quad,\quad
\Psi_2(t) = \Psi_2\,\exp\Big( - \frac{i}{\hbar}\,E_2t\Big),
\end{eqnarray}
where $E_1$ and $E_2$ are the energies of muonium in the ground and
excited states, then $\Gamma_{M^0} = \hbar/\tau_{M^0}$ with $\tau_{M^0}
\sim 10^6\,sec$ is the mean life of muonium.

\begin{figure}
 \includegraphics[width=150pt]{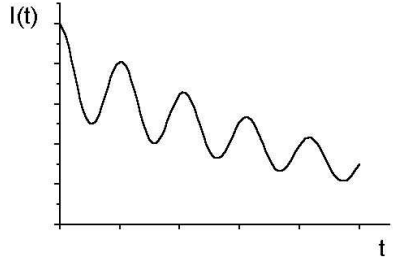}
 \centering\caption{The decay curve of muonium with account for the excited
 state.} \label{f:3}
\end{figure}

Taking into account that the wave function of muonium is normalized to
the current density and introducing the power of the polarization
$\varepsilon = |C_2|^2/|C_1 + C_2|^2$, it is not difficult to
calculate the current density of $M^0$ at an arbitrary time $t$ [7]:
\begin{eqnarray}\label{label3.4}
I(t) = I_0\,\left\{1 -2\,\sqrt{\varepsilon}\,(1 -
\sqrt{\varepsilon})\left[1 - \cos\left(\frac{E_2 -
E_1}{\hbar}\,t\right)\right]\right\}\,
\exp\left(-\frac{\Gamma_{M^0}t}{\hbar}\right).
\end{eqnarray}
It is seen that the decay curve $I(t)$ depicted in Fig.\ref{f:3} contains
oscillations. Let us estimate the period of oscillations (the distance
on which the curve in Fig.\ref{f:3} contains only one oscillation in
time). Using (\ref{label3.4}), denoting $\omega = (E_2 - E_1)/\hbar =
2\pi/T$ and accounting for that $E_2 - E_1 =10.1\,{\rm eV}$ one obtains
$T\approx 4\times 10^{-16}\,$s. When matching $T$ with $\tau_{M^0}$
one can see that the period of oscillations in time is smaller
compared with $\tau_{M^0}$ by ten orders of magnitude. The former
means that in Fig.\ref{f:3} there should be a lot of oscillations for the mean
life of $M^0$.

Since muonium produces in the reaction (\ref{label2.1}) with $E_{M^0}
\gg m_{M^0}c^2$, for example, for Tevatron--DIS one gets $E_{M^0} \sim
225\,{\rm MeV}$, so $T_{\ell ab} = T(E_{M^0}/m_{M^0}c^2) \approx
10^{-12}\,{\rm s}$. For this time muonium should be moved for the
distance comeasurable with a space period of oscillations equal to
approximately 300$\,mkm$. Effects on such distances are measurable.

Let us estimate the value of $\varepsilon$ defining the contribution
of the wave function $\Psi_2$ to the formula (\ref{label3.2}) due to
the production of muonium at the excited state $M^{0*}$. There are two
mechanisms of the production of $M^{0*}$: (i) the production at the
vertex of recombination $e^- + \mu^+ \to M^{0*}$ in the diagrams in
Fig.\ref{f:1} and (ii) interaction in the final state when $M^0$ rescatters
inelastically by either $\mu^-$ or the nucleus. Consider the first
mechanism. Let $M^0$ be produced at the $2\,{^1}{\rm S}_0$--state. In
this case the equality (\ref{label2.3}) would contain
$\Phi(\vec{p}\,)$ defined by
\begin{eqnarray}\label{label3.5}
\Phi(\vec{p}\,) = - \frac{32\sqrt{2\pi a^3_0}\,(1 -
4\,\vec{p}^{\,2}\,a^2_0)}{(1 + 4\,\vec{p}^{\,2}\,a^2_0)^3},
\end{eqnarray}
where $a_0$ has been determined above and $\Psi(0) = 1/\sqrt{8\pi
a^3_0}$. This leads to the value $\varepsilon = 1/256$ which means
that the admixture of the excited state $\Psi_2$ in
Eq.(\ref{label3.2}) is of order $10\,\%$. In the case of the $M^{0*}$
production at one of the $2P$--states the value of $\varepsilon$ turns
out to be of the same order. The second mechanism gives a contribution
substantially less.

Thus, a comparison of the decay curves of muonium and antimuonium for
a few periods of oscillations should give a possibility to check
whether there exists CPT--invariance for bound states in QED.

\end{document}